# TLR-exosomes exhibit distinct kinetics and effector function

Short Title: Differential effector function of TLR-exosomes

Swetha Srinivasan, Michelle Su, Shashidhar Ravishankar, James Moore, PamelaSara E Head, J. Brandon Dixon, Fredrik O Vannberg*

**Affiliations**

**School of Biology, Georgia Institute of Technology, Atlanta, GA, USA**

Swetha Srinivasan, James Moore, Michelle Su, PamelaSara Head, Shashidhar Ravishankar, Fredrik Vannberg*

**Parker H. Petit Institute for Bioengineering and Bioscience Georgia Institute of Technology**

Swetha Srinivasan, J. Brandon Dixon, Fredrik Vannberg*

**Woodruff School of Mechanical Engineering, Georgia Institute of Technology, Atlanta**

J.Brandon Dixon

**Current address: Emory University Graduate Division of Biological and Biomedical Sciences Atlanta, GA 30322, USA.**

PamelaSara Head, Michelle Su

**Materials and correspondence**
Dr Fredrik Vannberg
IBB 3312
315 Ferst Dr.
Atlanta, GA 30332-0405
Phone: 404-645-5964
Fax: 404-894-2291
vannberg@gatech.edu

**Word Count: 3005**

**Figure Count: 6**

# ABSTRACT

The innate immune system is vital to rapidly responding to pathogens and Toll-like receptors (TLRs) are a critical component of this response. Nanovesicular exosomes play a role in immunity, but to date their exact contribution to the dissemination of the TLR response is unknown. Here we show that exosomes from TLR stimulated cells (TLR-exosomes) can largely recapitulate TLR activation in distal cells in vitro. We can abrogate the action-at-a-distance signaling of exosomes by UV irradiation, demonstrating that RNA is crucial for their effector function. We are the first to show that exosomes derived from poly(I:C) stimulated cells induce in vivo macrophage M1-like polarization within murine lymph nodes. These TLR-exosomes demonstrate enhanced trafficking to the node and preferentially recruit neutrophils as compared to control-exosomes. This work definitively establishes the differential effector function for TLR-exosomes in communicating the activation state of the cell of origin.

# Introduction

Detecting microbial pathogens rapidly and containing their spread is a critical function of the innate immune system (1). Toll-like receptors (TLR) are an essential arm of innate immunity as they detect highly conserved pathogen associated molecular patterns (PAMPs) and play an important role in host cell defense (2). The direct response of cells stimulated TLR agonists locally is well characterized (3, 4). Dendritic cells when exposed to the TLR4 agonist lipopolysaccharide (LPS) show a distinct gene expression response as compared with cells exposed to the TLR3 agonist poly I:C (pIC), and these gene expression profiles are known to be pathogen specific (5). TLR stimulation induces production of a broad range of molecules including cytokines and chemokines (6), which are essential for host response to infection as well as for the development of an adaptive immune response (7).

While cytokines and chemokines are well studied for their roles in mediating cell-cell communication to establish immunity (8), recently more complex messengers such as extracellular vesicles have been discovered (9). Exosomes are nanovesicles (30-150nm in diameter) released by the fusion of large multivesicular endosomes with the host cell membrane (10) . They are released by most known cell types, ubiquitously found in biological fluids (11) and carry functional cargo in the form of mRNA, miRNA and proteins to distal recipient cells where the contents can modulate the recipient cell phenotype (12).

Exosomes have many distinct roles in physiology and immunity (13-15) and have known to play dual roles in both immune system activation (16) and immune suppression (17). Furthermore, we recently showed that exosomes are rapidly trafficked from peripheral tissues by the lymphatics, and retained in the draining lymph node by macrophages in a murine model (18).

While local cellular response to TLR stimulation is well studied both *in vitro* and *in vivo* (19-21), the role of exosomes in the distal dissemination of the TLR response is less well understood. We speculated that exosomes from TLR stimulated cells could potentially transmit information to distal unexposed cells *in vitro*. Moreover, we wanted to understand the impact of stable lymphatic retention of exosomes by macrophages in the development of an immune response *in vivo*.

Here we delineate the differential effector function of exosomes based on the innate immune activation state (control-, LPS-, poly(I:C)-stimulation) of the cell of origin in both *in vitro* and *in vivo* experiments. Poly(I:C) derived exosomes are rapidly transported to the lymph node and polarize distal macrophages to an M1-like state and recruit neutrophils. We show that exosomes are reprogrammed to carry a TLR-specific message to distal cells and more work is warranted to understand the implications of this in immunity to pathogens and cancer.

## Results
### Characterization of exosomes
To understand the effect of exosomal cargo released from locally stimulated cells on distal cell expression, we collected exosomes from local ovarian adenocarcinoma (HEY) cells that were either unstimulated (control exosomes), or stimulated with poly(I:C) (pIC exosomes), or lipopolysaccharide (LPS exosomes) for 48 hours. The three groups of exosomes were added to naïve (distal) cells and the changes in gene expression profiles were compared between local TLR stimulation (for 6 hours) and distal stimulation mediated by exosomes (for 48 hours) on a microarray (**Fig. 1a**).

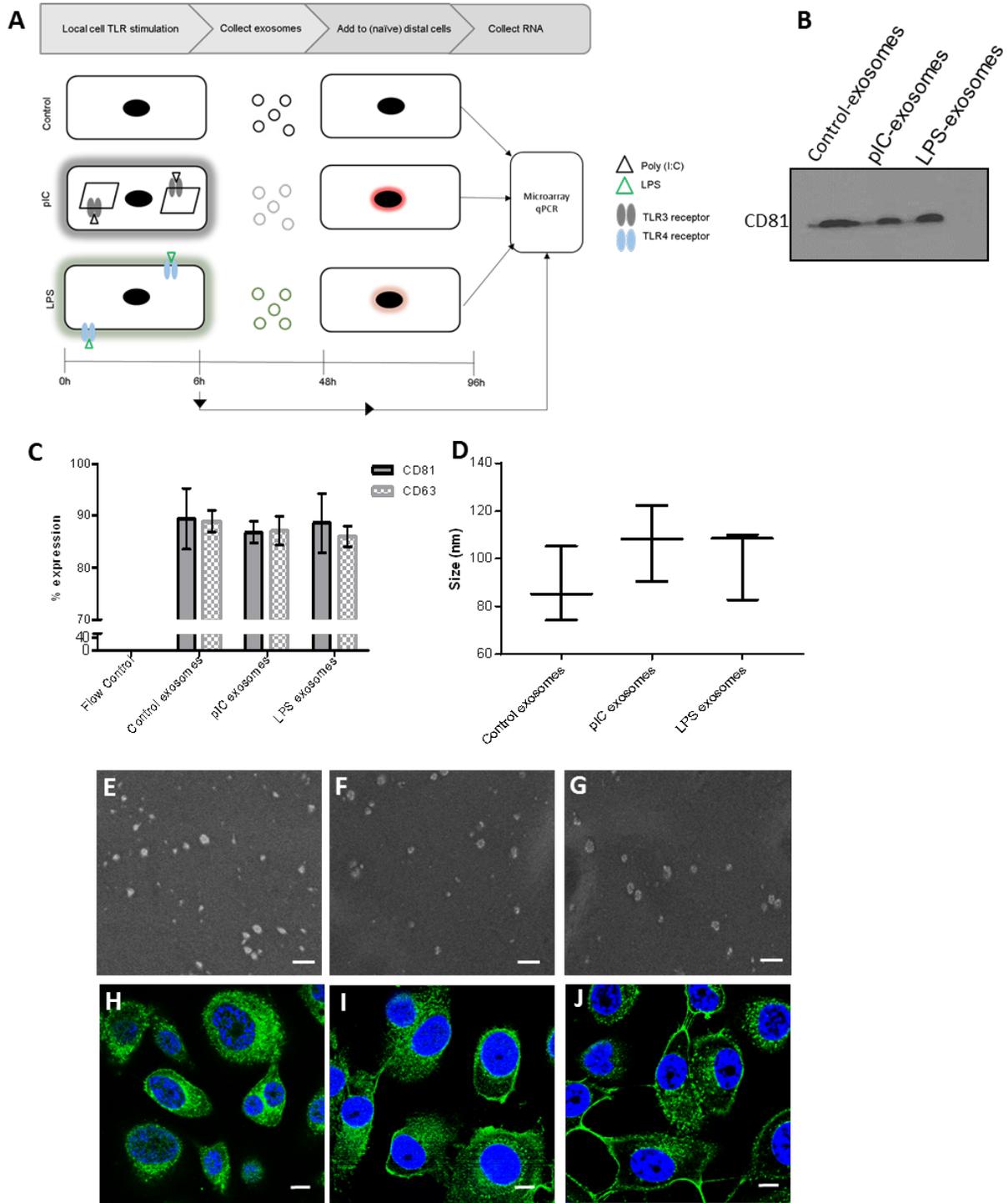

**Figure 1:** Experimental setup, biophysical and biochemical characterization of exosomes. (**a**): Schematic showing experimental setup where local cells are either unstimulated or directly stimulated with TLR agonists LPS and poly(I:C) for 6 hours. Exosomes are collected from the local cells and exposed to distal cells for 24 hours and then both local and distal cells are profiled by microarrays (**b**) Expression of CD81 and CD63 on the surface of control, pIC and LPS exosomes as quantified by flow cytometry (**c**) Size distribution profiles of control, pIC and LPS exosomes quantified on a Zetasizer (**d**) Western blot of CD81 protein expression on control, pIC and LPS exosomes. Scanning electron micrographs of (**e**) Control

exosomes, (**f**) pIC exosomes and (**g**) LPS exosomes showing characteristic spherical shape. Scale bars, 500 nm. Confocal images of distal cells showing uptake of PKH67 labeled (**h**) Control exosomes, (**i**) pIC exosomes and (**j**) LPS exosomes. Scale bars, 10 µm.

Nanovesicular exosomes were obtained through ultracentrifugation as described previously (18) and were found to have the canonical exosomal tetraspanins CD81 and CD63. We found comparable surface CD81 protein expression by western blot across the control, pIC and LPS exosomes (**Fig. 1b, Supplementary Fig. 1a**) and flow cytometry showed comparable levels of expression of CD81 and CD63 on control, pIC and LPS exosomes (**Fig. 1c, Supplementary Fig. 1 b-c**). The size distribution profiles were similar for the three groups (**Fig. 1d**) and we noted that control-, pIC- and LPS-exosomes was spherical, as revealed by scanning electron micrographs (**Fig. 1 e-g respectively)**. Finally, the uptake of control-, pIC- and LPS-exosomes by distal cells was comparable (**Fig. 1 h-j**). Therefore, the three groups of exosomes did not differ in biophysical and biochemical properties or cellular uptake.

## Effect of LPS on local and distal cells

We analyzed microarray data to identify genes that were differentially expressed in i) local and ii) distal cells with respect to unstimulated cells. Our findings indicate that local cells stimulated with LPS as well as distal cells stimulated with LPS-exosomes both show enhanced expression of key inflammatory genes **(Fig. 2a)**. The classical LPS response occurs via binding to TLR4 which results in activation of an inflammatory response and release of TNFA, CCL3 and IL1B. The cell then enters a refractory state resistant to further LPS stimulation characterized by increased *TOLLIP* expression and NF-κB inactivation (22) (**Fig. 2b**).

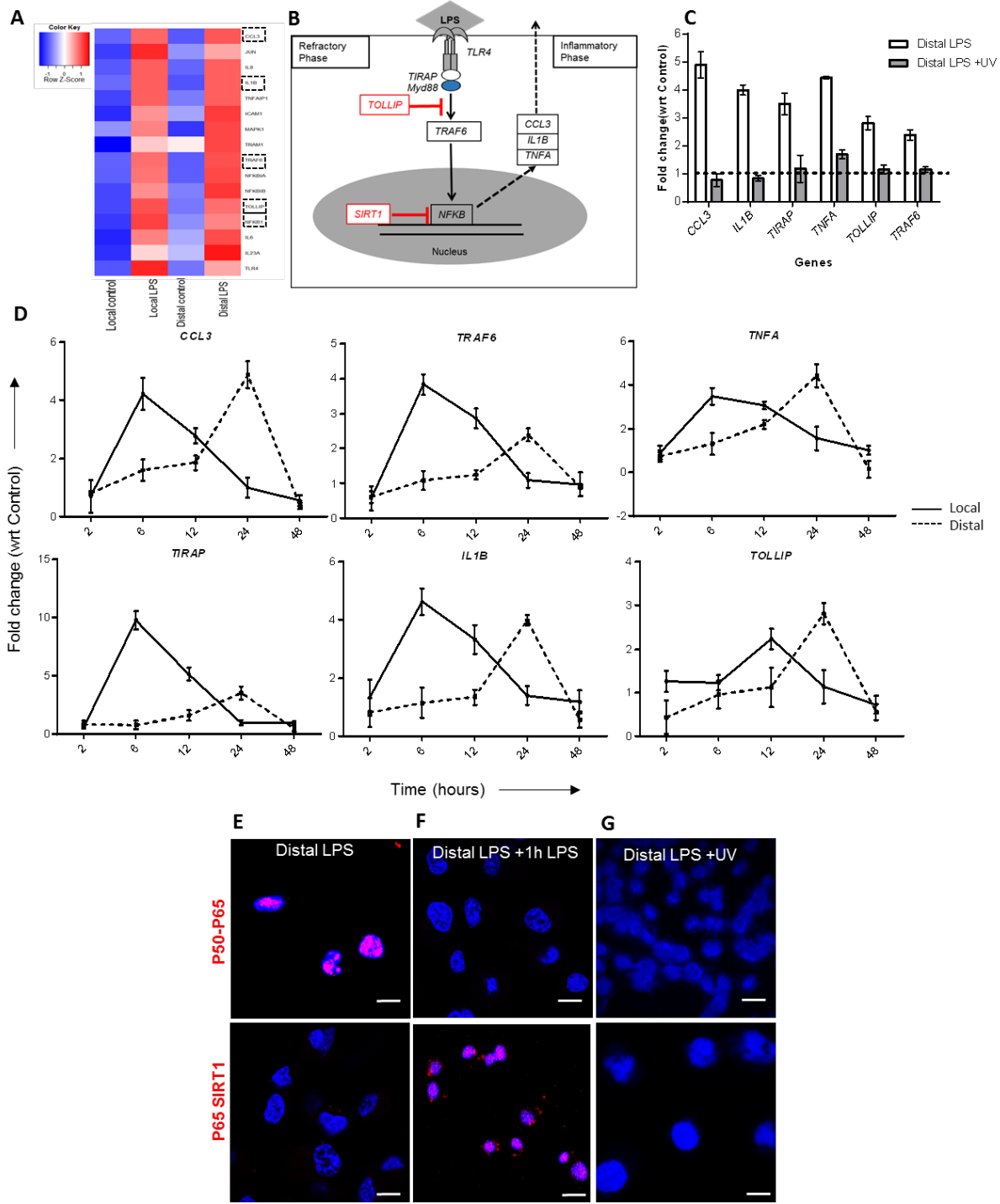

**Figure 2:** The LPS response in local and distal cells. (**a**) Heatmap of selected inflammatory genes involved in LPS response in local and distal cells. (**b**) Classical LPS response pathway showing the key genes and inhibitors that establish the initial inflammatory phase and the subsequent refractory phase. (**c**) Distal cell gene expression after exposure to either LPS exosomes or UV irradiated LPS exosomes (24 hours post exosome addition). (**d**) Time course of gene expression comparing local cell response to LPS (solid line) against distal cell response to LPS exosomes (dotted line) for the genes shown. Proximity ligation

assay showing (**e**) NF-κB activation state in distal cells with LPS exosome stimulation as measured by colocalization of p50-p65 subunits, (**f**) SIRT1 mediated inactivation of NF-κB after restimulating distal cells with 1 hour of LPS as measured by colocalization of NF-κB subunit p65 with SIRT1 colocalization and (**g**) inactivation of NF-κB by pretreating LPS exosomes with UV. Scale bars, 20μm.

Exposure to LPS exosomes caused distal cells to upregulate molecular pathways associated with chemokine signaling as well as TLR activation (**Supplementary Fig. 4a**). To examine the temporal response of local cells to LPS and distal cells to LPS-exosomes we evaluated changes in gene expression with respect to control cells using qRT-PCR (**Fig. 2d**). We looked at *CCL3, IL1B, TNFA, TIRAP, TRAF6* and *TOLLIP* expression over 48 hours and observed that peak local cell response to LPS stimulation was at 6 hours with *TOLLIP* expression peaking at 12 hours. Distal cells responded to LPS-exosome stimulation at 24 hours to levels comparable with local cells.

Previous studies have utilized UV light to degrade the RNA payload of exosomes, and we see similar UV-based elimination of exosomal RNA as detected by chip-based electrophoresis. (**Supplementary Fig. 5 a-c**). Distal cells exposed to UV treated LPS-exosomes showed significantly decreased expression of the six gene panel as compared with LPS-exosomes, and recapitulated the response of cells exposed to control-exosomes. This indicates that UV irradiation, likely through degradation of RNA, was necessary and sufficient to modulate the effector function of exosomes in distal cells (**Fig. 2c**).

LPS-exosomes contained ~4% carryover of initial LPS dose (**Supplementary Fig. 2a**). We exposed local cells to this carryover dose and examined gene expression at 6h, 12h and 24h time points (**Supplementary Fig. 2c,d**) which shows limited contribution to distal response.

A proximity ligation assay (PLA) was used to determine the co-localization of the p50 and p65 subunits of NF-κB to indicate activation of this complex. Local cells exposed to LPS and distal cells treated LPS exosomes showed co-localization of p50 and p65 subunits indicating that the NF-κB complex was active in both (**Supplementary Fig. 3a, Fig. 2e**).

Endotoxin tolerance is a protective mechanism to prevent overt inflammation in response to LPS and results in a transient unresponsive cellular state, characterized by the epigenetic inactivation of the p65 subunit of NF-κB by the histone deacetylase SIRT1 (23). Local and distal cell response to LPS re-stimulation, and resulting impact on NF-κB silencing state, was analyzed using PLA (**Supplementary Fig. 3b**). We exposed distal cells to LPS-exosomes and then re-stimulated them with LPS and we found that these cells had hallmarks of endotoxin tolerance with p65-SIRT1 co-localization resulting in NF-κB inactivation (**Fig. 2f**). Additionally, UV irradiation of exosomes abrogated i) NF-κB activation and ii) *SIRT1* mediated inactivation in distal cells (**Fig. 2g**).

### Effect of pIC on local and distal cells

We analyzed microarray data to detect genes involved in i) local pIC and ii) distal pIC-exosomes responses as compared with unstimulated cells, in particular with respect to genes involved in antiviral activity (**Fig. 3a**). The classical pIC response occurs via endosomally localized TLR3 binding pIC, or cytoplasmic MDA5 resulting in NF-κB activation and interferon production[2] (**Fig. 3b**).

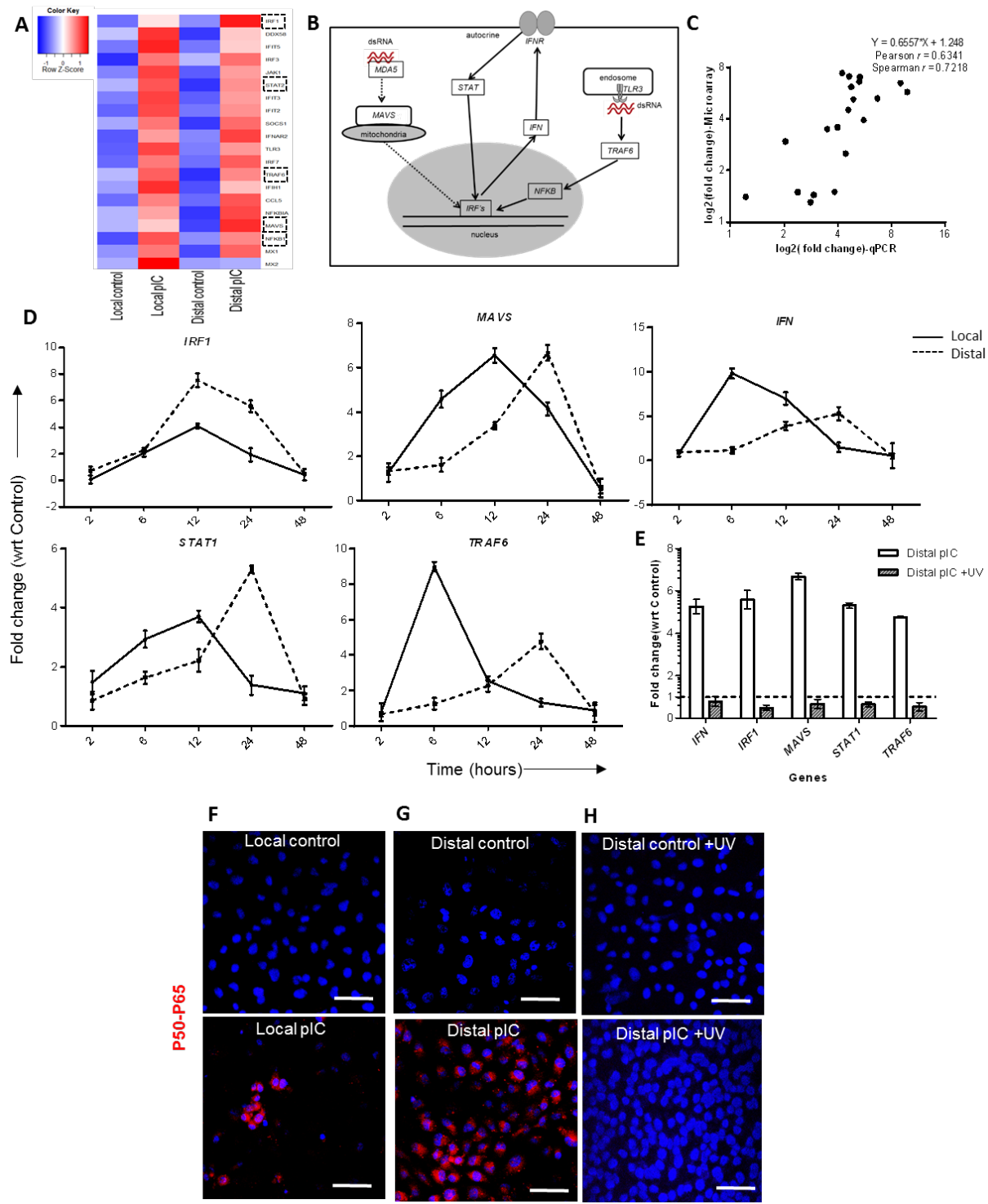

**Figure 3:** The pIC response in local and distal cells. (**a**) Heatmap of selected antiviral genes involved in viral RNA response in local and distal cells. (**b**) Pathways of intracellular response to pIC showing both the cytoplasmic recognition pathway on the left and the endosomal pathway on the right. (**c**) Scatter plots showing the correlation between the fold change detected via qPCR when compared to microarrays (**d**) Time course of gene expression comparing local cell response to pIC (solid lines) against distal cell response to pIC exosomes (dotted lines) for the genes shown. (**e**) Distal cell gene expression after exposure to either pIC exosomes or UV irradiated pIC exosomes (24 hours post exosome addition). Verifying NF-κB activation in (**f**) local cells with pIC, (**g**) distal cells with pIC exosomes or (**h**) distal cells with UV treated pIC exosomes confirming the colocalization of P50-P65 subunits using a proximity ligation assay. Scale bars, 50 μm.

Distal cells exposed to pIC-exosomes were found to upregulate the following pathways: *TLR/RXR activation* and *RIG-1 like receptor* (**Supplementary Fig. 4b**). To examine the temporal response of local and distal cells to pIC/pIC-exosomes, respectively, we evaluated changes in gene expression using qRT-PCR (**Fig. 3d**). Analyzing *IRF1, MAVS, IFN, STAT1* and *TRAF6* gene expression over 48 hours we observed that local cells responded to pIC stimulation between 6-12 hours, and that distal cells responded to pIC-exosome stimulation between 12-24 hours at comparable levels. Additionally, we found high concordance between local and distal cell gene expression using both microarrays and qRT-PCR (**Fig. 3c**).

We could not detect any carryover of pIC to distal cells by pIC-exosomes (**Supplementary Fig. 2b**). Distal gene expression after exposure to UV-irradiated pIC-exosomes showed a complete abrogation of changes in gene expression as detected by qRT-PCR (**Fig. 3e**). Exposing local cells to pIC and distal cells exposed to pIC-exosomes lead to NF-κB activation as detected by PLA signal of the active P50-P65 heterodimer (**Fig. 3f, g** respectively), which was abrogated by UV-irradiation of pIC-exosomes (**Fig. 3h**).

### Effect of exosome uptake on macrophages

To understand the impact of exosome uptake on macrophage function *in vivo,* mice were injected with either PBS, control-exosomes or pIC-exosomes. We verified the uptake of control- and pIC-exosomes by macrophages in murine lymph nodes using co-localization of fluorescently labeled exosomes and CD11b (**Supplementary Fig. 7**). Lymph nodes were extracted at 48 hours post injection followed by isolation CD11b macrophages and subsequent analysis by RNA-Seq.

Whole lymph nodes were sectioned to verify protein expression by immunohistochemistry (**Fig. 4a**).

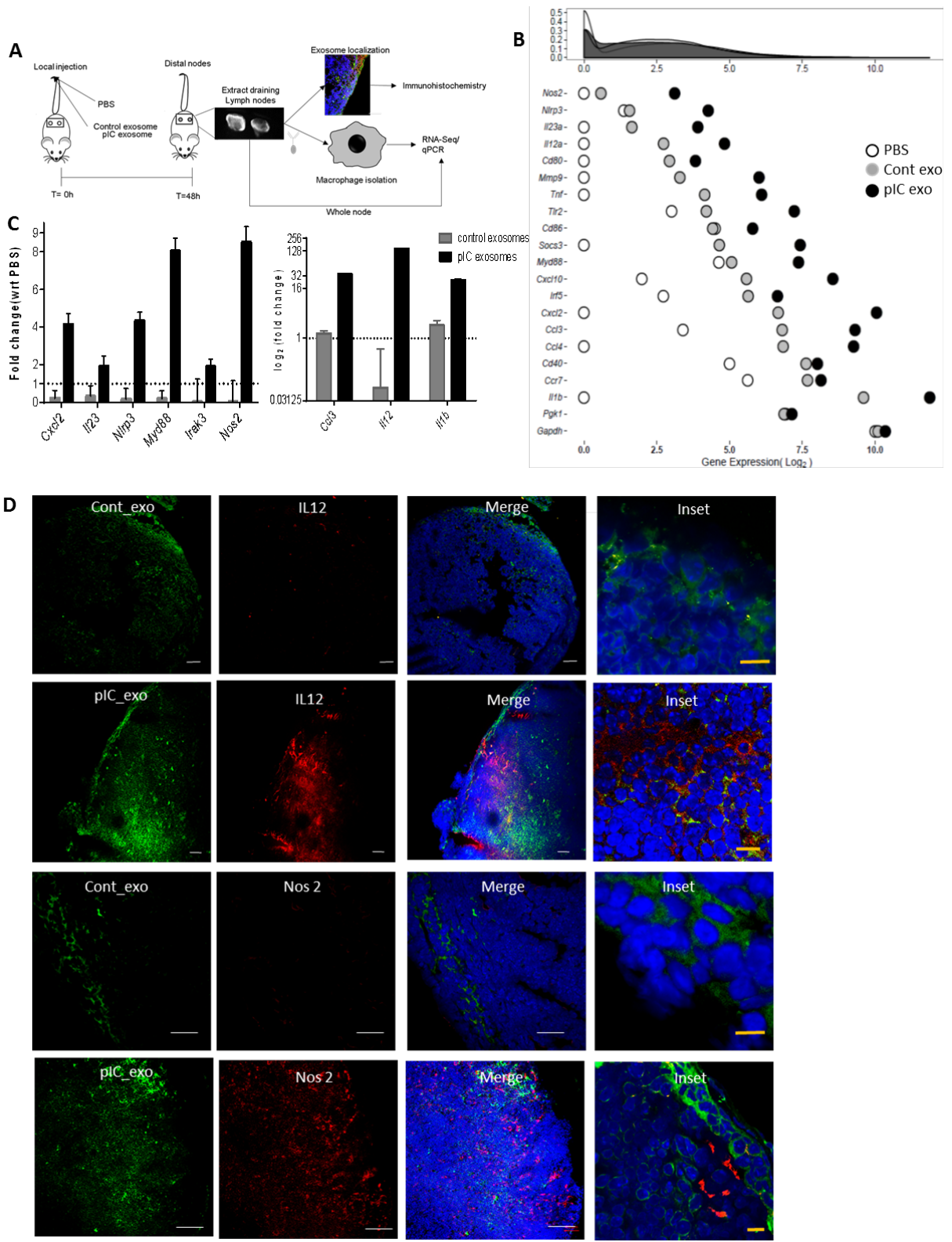

**Figure 4:** Effect of exosome uptake on macrophages. (**a**) Schematic showing mouse study with injection of either PBS, control-, or pIC-exosomes locally followed by excision of distal draining lymph nodes,

isolation of macrophages and analysis of macrophage gene expression by next generation sequencing. (**b**) Relative expression of key M1 and housekeeping genes in macrophages after exposure to PBS, control- or pIC-exosomes. (**c**) Validation of key M1 genes in macrophages with control- or pIC-exosomes as compared to PBS by qRT-PCR. (**d**) Validation of M1 markers Nos2 and Il12 expression in lymph node sections after exposure to control- or pIC-exosomes. White scale bars, 50 μm, yellow scale bars, 10 μm.

The relative gene expression of macrophages with PBS, control- and pIC-exosomes is depicted for selected genes (**Fig. 4b**). pIC-exosomes strongly polarize the macrophages to an M1 like state with a significant shift in gene expression, while control exosomes show an intermediate phenotype (**Fig. 4b**). Macrophages with pIC-exosomes *in vivo* show an enrichment of pro-inflammatory pathways including *NF-κB signaling* and *Chemokine signaling pathways* (**Supplementary Fig. 6a**). A complete list of genes enriched in pIC-exosomes with respect to PBS and control exosomes is provided (**Supplementary Table 3-4**).

The gene expression of M1 markers was validated using qRT-PCR (**Fig. 4c**). We find that macrophages from mice with pIC-exosome injection had a significantly higher expression of all the selected M1 markers when compared to control-exosomes (**Fig. 4c**). The correlation between macrophage gene expression with PBS, control- or pIC-exosomes determined by RNA-Seq and validation by qRT-PCR was high (**Supplementary Fig. 6b**). Whole lymph node sections with pIC exosomes showed high expression of *Il12, Nos2, Cd86,* and *MhcII*, while sections with control-exosomes showed low to no expression of the same markers (**Fig. 4d**, **Supplementary Fig. 8**)

## Impact of lymphatic trafficking of exosomes on draining lymph node retention

To investigate differences in lymphatic transport of control- and pIC-exosomes, mice tails were imaged using the NIR system described previously[18]. We observed that the kinetics of transport of pIC exosomes in both the dominant and non-dominant vessel was considerably higher than that of control-exosomes in both lymphatic vessels (**Fig. 5a**, **Supplementary Fig. 9a-h**).

Similarly, pIC-exosomes were retained to a much higher extent in the draining lymph nodes as compared to control-exosomes, a difference that can be seen as early as five minutes after intradermal injection (**Fig. 5b**, **Supplementary Fig. 9i-l**). The transport times and packet frequencies of control- and pIC-exosomes were not significantly different in the vessels or lymph nodes (**Supplementary Fig. 10**). The total packet transport, which is a measure of the indicative of total pump flow that takes into account both the amplitude of lymphatic contraction and the frequency of pumping, is significantly higher for pIC-exosomes than control exosomes Taken together, this data suggests that exosomes released from pIC stimulated cells enhance lymph flow and the subsequent retention of exosomes at the lymph node (**Fig. 5c**).

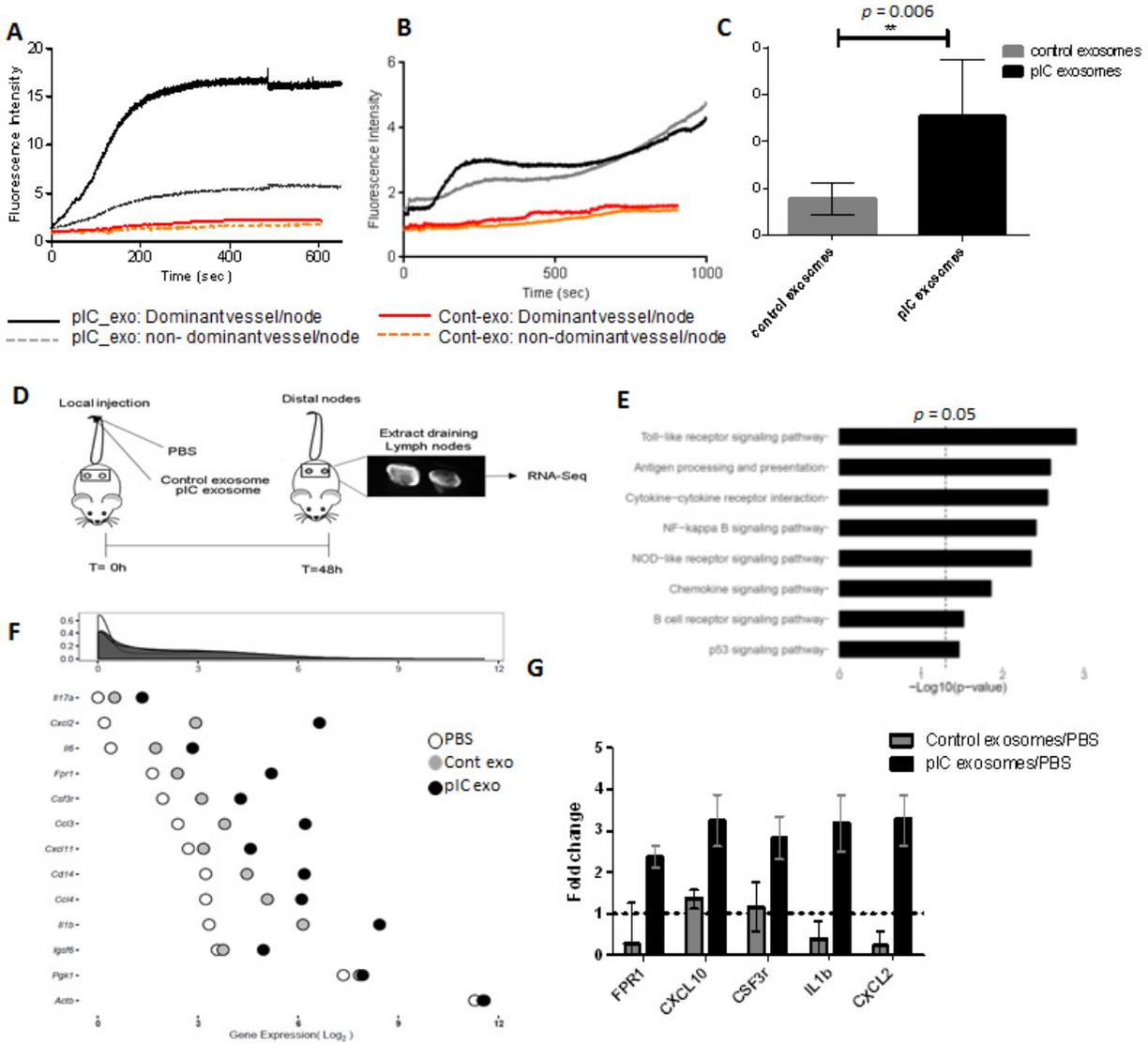

Fig. 5: Impact of lymphatic trafficking of exosomes on draining lymph node retention (a) Kinetics of lymphatic vessel transport comparing control and pIC exosome trafficking in the dominant and non-dominant vessels. (b) Comparison of kinetics of draining lymph node retention of control and pIC exosomes in dominant and non-dominant nodes. (c) Lymphatic packet transport of control and pIC exosomes. (d) Schematic of experiment showing PBS, control or pIC exosomes in mouse tail, followed by lymph node extraction at 48 hours followed by RNA-Seq of whole lymph nodes. (e) Pathway analysis showing pathways enriched in pIC exosomes in whole nodes with respect to PBS and (f) Relative expression of key neutrophil markers in whole lymph nodes after exposure to PBS, control or pIC exosomes. (g) Validation of neutrophil recruitment and inflammation in the whole node with control or pIC exosomes as compared to PBS by qPCR.

To understand the impact of exosome retention on the lymph node *in vivo,* the lymph nodes were extracted at 48 hours post exosome injection and analyzed by RNA-Seq (**Fig. 5d**, **Supplementary Table 5**). Pathway analysis of whole nodes comparing pIC-exosomes with PBS shows an increase in pro-inflammatory signaling pathways (**Fig. 5e**). We extracted the gene expression of the 546 genes utilized to estimate the abundance of 22 immune subsets based on the CIBERSORT algorithm, extending the algorithm to utilize our RNA-Seq data (24). Using this analysis we found evidence of neutrophil, mast cell and monocyte recruitment in the pIC-exosomes group as compared to the PBS and control-exosome groups. The relative gene expression of whole nodes with PBS, control- and pIC-exosomes for selected neutrophil markers is shown in **Fig. 5f**. The gene expression of neutrophil recruitment markers and inflammatory signals in the whole node was validated using qRT-PCR (**Fig. 5g**). We find that whole lyph nodes from mice with pIC-exosome injection had a significantly higher expression of all the selected genes when compared to control-exosomes (**Fig. 5g**)

Whole lymph node sections with pIC exosomes showed high expression of *Gr1* and *Ly6g* while sections with control-exosomes showed low to no expression of the same markers (**Fig. 6a-b**).

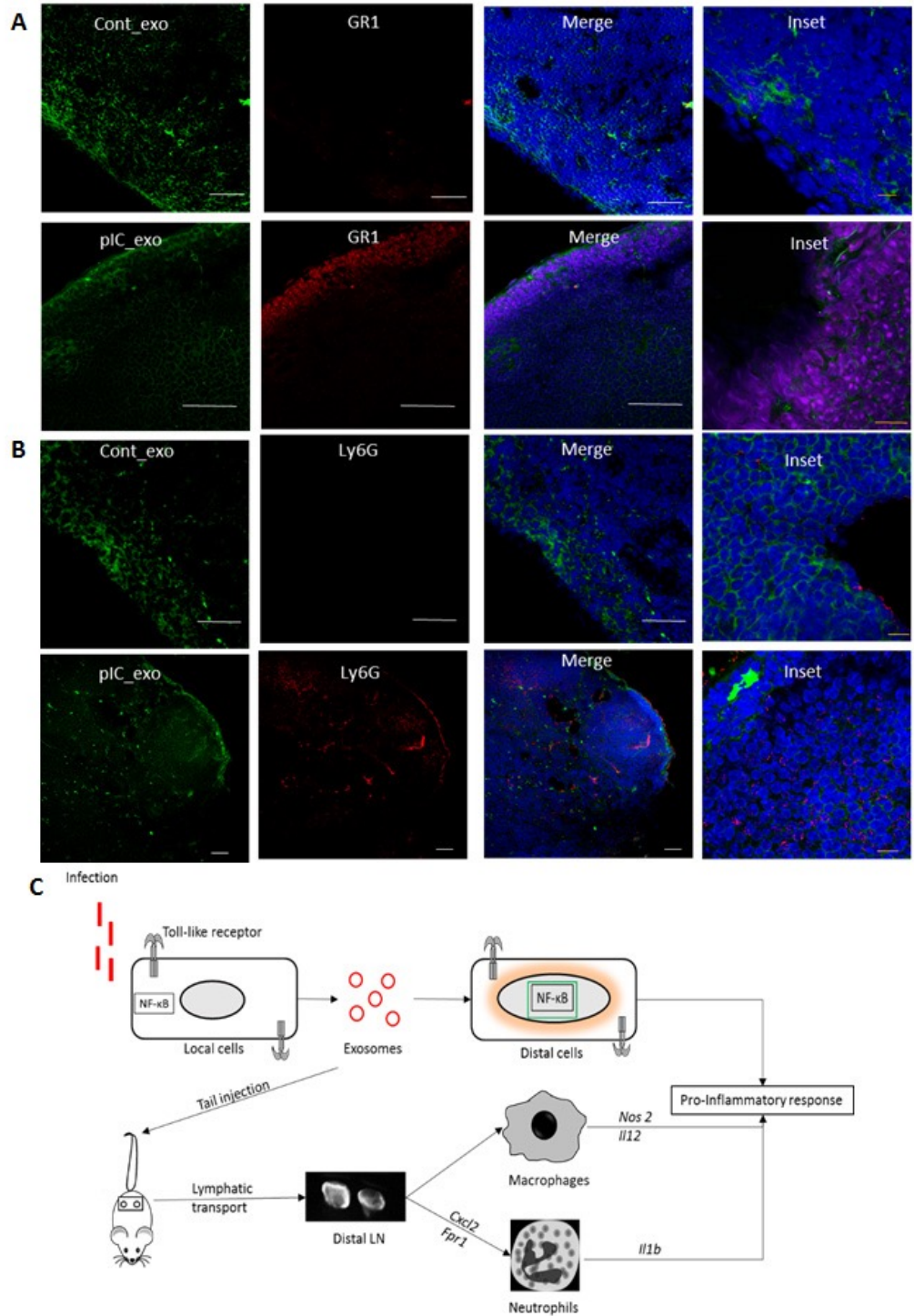

Figure 6: Validation of neutrophil recruitment to node. A) Immunohistochemistry of lymph node sections showing increase of neutrophil markers after pIC exosome uptake A) GR1 expression and B) Ly6G expression after control and pIC exosome uptake. C) Model of exosome action showing transmittance of local cell TLR activation to distal cells resulting in a pro-inflammatory response both in vitro and in vivo.

Finally, our model for exosome mediated cell-cell communication in the dissemination of the TLR response is shown (**Fig. 6c**). We have shown that both LPS- and pIC-exosomes elicit a pro-inflammatory or antiviral gene response in vitro that is characterized by the activation and localization of NF-κB in the nucleus. Furthermore, exosomes are taken up by macrophages and polarized to a M1-like pro-inflammatory state characterized by the increased expression of *Il12* and *Nos2* while the node is reprogrammed to recruit neutrophils and shows increased inflammation.

## Discussion

The ability of the immune system to recognize and respond to foreign organisms is essential to survival and TLR's play a central role in this response (6). Exosomes have many distinct roles that vary depending on their cell of origin, from modulating the immune response(13), to cancer progression and immune evasion (14). Here, for the first time we show that distal target cells are primed to respond to imminent pathogen attack via TLR mediated NF-κB activation and the resulting secretion of pro-inflammatory cytokines.

Our experiments demonstrate reproducible perturbations of distal cells with minimal to nonexistent agonist carryover demonstrating the importance of the immune state of the cell of origin. We observe a ~12-18 hour shift in distal cell response likely corresponding to the time required for exosome uptake, unpacking and release of contained biomolecules within the distal cells. Similar timing has been noted for luciferase mRNA delivery and subsequent expression within a glioblastoma exosome delivery model (25). Additionally, exosomal RNA was able to recapitulate the inflammation and apoptosis in cells post *M. tuberculosis* infection (26) We, like others before us, suggest that RNA is likely the key contributor to the effects observed since the relative dose of UV light in this experiment completely degrades RNA but has minimal potential to degrade/inactivate protein (27, 28).

The discovery of pathogen derived components from infected cells and the subsequent effect on immune activation is well studied (29). Here we show recapitulation of infection response via exosomes from TLR stimulated cells underscoring the contribution of the host cell to the pathogen response. Furthermore, we provide a model to delineate the contribution of the host and pathogen to the response elicited in distal cells post exosomes exposure.

We also show that exosomes derived from TLR stimulated cells epigenetically modify distal cells to be refractory to further LPS stimulation (i.e. undergo endotoxin tolerance), thus protecting these distal cells from uncontrolled inflammation (23). While the TLR response and endotoxin tolerance is well characterized in immune cells, it is less understood in non-immune cells (2). To our knowledge this is the first study to demonstrate that exosomes from non-immune cell origin can i) epigenetically modify and ii) convey TLR-specific information to downstream cells. Further studies are needed to understand the contribution of exosomes from non-immune cells to the innate and adaptive immune response.

The biological importance of exosomes in the crosstalk between cancer cells and the immune subsets is a growing area of research. Ovarian cancer is associated with one of the highest mortalities (30) and survival statistics have not improved significantly over the past three (31), highlighting the need to better understand the tumor microenvironment. Inflammation is a hallmark of *in situ* tumors and has important implications for cancer cell survival, proliferation and migration (32). TLR activation of the tumor microenvironment impacts tumor growth kinetics (33-35), and actively contributes to inflammation and tumor derived exosomes likely disseminate this information impacting cancer fate.

Administration of pIC in mouse models of melanoma, lung cancer and colon cancer elicited robust anti-tumor immune responses (36) and our present work suggests that pIC-exosomes derived from the site of pIC injection could be playing a role in this response. Researchers would be wise to consider the potential role of TLR-exosomes and their potential to impact on diverse cancer immunotherapy regiments.

Subcapsular sinus (SCS) macrophages play roles in suppressing the spread of melanoma thus enhancing immunity (37). We previously showed an important role for SCS macrophages in exosome retention (18) that we expand to demonstrate that ovarian cancer-derived exosomes are capable of inducing an inflammatory response via *Il12 and Nos2* in murine macrophages *in vivo*. *Nos2* results in nitric oxide (NO) production by macrophages contributing to diverse mechanisms including cytotoxic activity against viruses and bacteria (38), reduction in lymphatic contractions (39) and inhibition of ovarian cancer growth (40).

The ability of pIC-exosomes to rapidly modulate lymphatic function within minutes of injection, to increase transport and nodal retention is striking considering that careful attention was taken to ensure total exosome dose, total fluorescence, and total injected volume were the same between all measurement groups. Collecting lymphatics vessels are known to exhibit fast functional responses to their environment (41) through direct mechanical or biological stimuli. Thus the response observed here could be the result of the direct interaction of exosomes with lymphatics, interactions with the immune cells that reside within the lymphatic wall (42) or through an indirect effect of the exosomes at the injection site on lymph formation. or through an indirect effect of the exosomes at the injection site on lymph formation. Furthermore, the enhanced NO signaling seen in lymph node resident immune cells 48 hours later, suggests that exosomes could continue flow modulation via immune cell mediated NO release, as has been shown in other inflammatory models (39). The data shown here is the first evidence that exosomes could directly be involved in modulating flow to the lymph node, a significant finding since the enhanced lymphatic function also leads to enhanced uptake of the exosomes within the cells of the draining lymph node. Given the reliance of exosomes and immune cell trafficking on lymphatic transport, the impact of exosome-mediated flow modulation, either through NO secretion by macrophages or some other unidentified mechanism, on the subsequent immune response warrants further study.

Neutrophils release cytokines and chemokines to coordinate the innate and adaptive immune responses and play active roles in antigen presentation (43). We saw an enrichment of neutrophil recruitment signals including FPR1 and CXCR2 (44) in the whole node accompanied with an increase in pro-inflammatory cytokines, such as IL1B. Thus, the lymph node microenvironment is being reprogrammed by exosomes to respond to the signals sent by the parent cells.

The ability of exosomes to carry siRNA, drugs, proteins and other molecules makes them ideal therapeutic vehicles (45). Several promising clinical trials for exosome based therapeutics were unable to get reproducible results in patients and we show that minor perturbations to the biological state of the cell of origin can contribute to exosome effector function. We show that TLR-exosomes have distinct and rapid kinetics, and that they can deliver fundamental innate immune signals faithfully. We show that UV irradiation can fundamentally reset the effector signal of pIC exosomes to control-exosomes, but it is not clear if this will reset the kinetics of lymph node transfer. Our work points out the potential of TLR-exosomes for therapeutic use, but it is also a cautionary tale of the how exosomes kinetics and effector function are a definite product of the physiological state of the cell of origin.

# Materials and methods

## Cell culture and TLR stimulation

Fetal bovine serum (Atlanta Biologicals, Lawrenceville, GA) was centrifuged for 15 hours at 120,000 g, 4$^O$C to remove exosomes and was used to make exosome free cell culture media. HEY cells (Cedarlane Labs, Ontario, Canada) were cultured in RPMI 1640 (Mediatech, Manassas, VA) supplemented with 10% exosome free fetal bovine serum, 2 mM L-glutamine, 10 mM HEPES buffer (both from Mediatech), penicillin (100 U/ml), and streptomycin (100 μg/mL) (Thermo Fisher Scientific, Waltham, MA) for 48 hours and the culture media was used for isolation of exosomes by ultracentrifugation. Ultra-pure *E.coli* K12 LPS and poly(I:C) ( Invivogen, San Diego, CA) were used to treat cells at concentrations of 100 ng/mL and 10 μg/mL respectively for most experiments. Fluorescent LPS- Alexa Fluor 594 (Thermo Fisher Scientific) and poly(I:C) Rhodamine (Invivogen) were used at the same concentration to determine exosome mediated carryover.

## Exosome isolation and characterization

Conditioned media was collected from HEY cells (with or without TLR agonist treatment) at 90% confluence for exosome isolation. Briefly, the culture media was spun at 300 g, for 10 minutes to remove dead cells followed by a spin at 16,500 g, 20 min. The supernatant was then filtered through 0.22 μm filters and centrifuged at 120,000 g for 120 min. The pellet containing exosomes was re-suspended in a suitable volume of PBS. The size homogeneity of vesicles obtained was checked using a Zetasizer Nano ZS90 (Malvern Instruments Ltd, Worcestershire, UK) and quantified using Pierce BCA Protein assay kit (Thermo Fisher Scientific).

## Flow Cytometry

To analyze the expression of exosomal surface markers, 4 μm aldehyde/sulfate latex beads (Thermo Fisher Scientific) were coated with anti-CD9 antibody (BD Biosciences, San Diego, CA; Cat: 555370) overnight and incubated with 30 μg of exosomes. The exosome-beads complexes were probed with Anti Human CD81-PE (BD Biosciences; Cat: 555676) or Anti human CD63-PE (BD Biosciences; Cat: 557305) and data was acquired on a LSR II Flow cytometer (BD Biosciences). Data analysis was performed using the FloJo software (FlowJo version 10, Ashland, OR).

## Confocal microscopy

The exosomes were labeled using PKH67 Green Fluorescent Cell Linker Kit for General Cell Membrane Labeling (Sigma-Aldrich, St. Louis, MO) as per the manufacturer's instructions. Briefly, exosomes in PBS were added to 500 μL of Diluent C and 2 μl of PKH67 dye was added to 500 μL of Diluent C. The two solutions were mixed and incubated for 5 min at room temperature. 1 ml of 1% BSA was added to stop the reaction. The labeled exosomes were centrifuged at 120,000 g for 70 min to remove excess dye. Labeled exosomes were added to 5*10$^5$ cell suspension, mixed gently for 2-3 min and seeded in 6-well plates. The cells were imaged after 24 and 48 hours using a Zeiss LSM 700 Image processing and data analysis were performed using the ZEN imaging software (Zeiss, Germany).

## Scanning electron microscopy

Exosomes were fixed with 3.7% glutaraldehyde (Sigma–Aldrich) on carbon stubs for 15 min. After washing twice with PBS, the fixed exosomes were dehydrated with an ascending sequence of ethanol (40%, 60%, 80%, 96–98%). After evaporation of ethanol, the samples were left to dry at room temperature for 24 h on a glass substrate, and then analyzed by Hitachi Cold Field Emission SEM SU8200 (Hitachi High-Tec, Tokyo, Japan).

### UV treatment of exosomes

Exosomes from control, and TLR agonist stimulated cells were re-suspended in PBS were then subjected to UV-light (254 nm) for 30 mins at 4°C to neutralize RNA carried within similar to previous studies(28).

### Nucleic acid extraction from exosomes

Exosomal RNA was extracted from exosome samples using the QIAamp viral RNA mini kit (Qiagen), and RNA was eluted with 40 μL buffer AVE, according to the manufacturer's instructions. RNA quality and quantity was analyzed using Nanodrop and Agilent Bioanalyzer chips

### Western blot

The total protein was extracted from cells and exosomes using modified RIPA buffer (Thermo Fisher Scientific) and cell debris was removed by centrifugation. Equal amounts of protein (15–20 μg) were then separated on polyacrylamide gels, transferred onto PVDF membranes (Bio-Rad, Hercules, CA, USA) and blotted using mouse anti-human CD81 Antibody (BD Biosciences; Cat: 555676). Membranes were developed using SuperSignal™ West Dura Extended Duration Substrate (Thermo Fisher Scientific).

### Microarray procedure and data analysis

Total RNA was isolated from control and TLR stimulated parental and recipient cells (grown with exosomes) after 48 hours using RNeasy mini RNA isolation kit (QIAGEN, Valencia, CA). The integrity of the RNA was verified using the Agilent 2100 Bioanalyzer (Agilent Technologies, Santa Clara, CA). mRNA's were converted to double stranded DNA and amplified using the Applause 3'-Amp System (NuGen, San Carlos, CA). This cDNA was fragmented and biotin labeled using the Encode Biotin Module (NuGen), hybridized to Affymetrix HG U133 Plus 2.0 oligonucleotide arrays and analyzed with a Gene Chip Scanner 3000 (Affymetrix, Santa Clara, CA). Raw data in the form of CEL files were produced by the Affymetrix GeneChip Operating System (GCOS) software.

mRNA microarray data were analyzed using the Expression Console software (Affymetrix) and Bioconductor tools(46) written in the R statistical programming language (www.rproject.org). Pre-processing of raw signal intensities and normalization was performed using GCRMA (R). Linear modelling of the transformed data was determined by using Limma(47) in R with the Benjamini and Hochberg correction. Differentially expressed probesets were identified using a threshold 5% FDR correction and a fold change ≥ 1.4 was applied. The microarray data has been uploaded to GEO (Gene Expression Omnibus, accession number GSE81248)

### Real-time quantitative PCR (qRT-PCR)

Parental HEY cells were grown with or without LPS or poly(I:C) stimulation and total RNA from the cells was collected at 2h, 6h, 12h , 24h or 48 hours. Similarly, control, LPS or poly(I:C) exosomes were added to recipient cells and total RNA was collected from cells grown with exosomes at 2h, 6h, 12h, 24h and 48 hours respectively. Total RNA from macrophages as well as parental and recipient cells was extracted using an RNeasy plus kit (QIAGEN), and cDNA was generated with SuperScript VILO cDNA Synthesis Kit (Thermo Fisher Scientific). Analysis was done on a Strategene Mx3005P System (Agilent Technologies) with SYBR Green PCR master Mix (Thermo Fisher Scientific). The (intron spanning) primers were used for quantitative real-time PCR are shown in Supplementary Tables 1 and 2. The fold change was calculated using the ΔΔCt method. All analyses were run in Prism 6 (GraphPad Software Inc, La Jolla, CA) and all data is presented as mean ± standard deviation.

## Proximity Ligation Assay (PLA)

HEY cells were cultured on 12-mm glass cell culture coverslips (Thermo Fisher Scientific). Cells were washed three times with phosphate-buffered saline (PBS) and then fixed for 15 min in PBS with 4% paraformaldehyde. After washing with gentle shaking, cells were permeabilized for 5 min with methanol and washed. The proximal-ligation assay to detect the interaction of p50 with p65, anti- NF-κB p50 (Santa Cruz Biotech, Dallas, TX; Cat: sc-8414) and anti-NF-κB p65 (Santa Cruz; Cat: sc-372) and for p65- SIRT1 interaction, anti-NF-κB p65(Santa Cruz Biotech; Cat: sc-8008) and anti-SIRT1 (Santa Cruz; Cat: sc-15404) were used with a Duolink PLA assay kit (Sigma Aldrich). Images were acquired Zeiss LSM 700 (Zeiss).

## Animal study and handling

Exosomes from unstimulated (Control) and poly I:C stimulated HEY cells were dual labeled with PKH67 and near infrared dye using IRDye® 800CW Protein labeling kit (Licor, Lincoln, NE). Control or Poly I:C exosomes (total quantity=10 μg) were injected intradermally into the tail of eight-week-old male Balb/C mice (Charles River Laboratories, Wilmington, MA) as described previously(18) and euthanized on day two( t=48 hours). PBS was mixed with PEGylated IRdye 800CW (Licor) and injected similarly as the experimental control. The LAL Chromogenic Endotoxin Quantitation Kit (Thermo Fisher Scientific) was used per the manufacturer's instruction to measure LPS concentration on all injected exosomes to ensure no endotoxin crossover. All procedures in this study have been approved by the Georgia Institute of Technology IACUC Review Board (Protocol #A15051).

## Lymph node extraction and macrophage isolation

The draining (sacral) lymph nodes, and control (axillary) lymph nodes were harvested from Control exosomes [Group 1; n=10], poly I:C exosomes[Group 2; n=10] and PBS control [Group 3; n=8]. Harvested lymph nodes from all groups were digested with collagenase D (Roche Ltd., Mannhein, Germany) and homogenized using 70 μm pore size strainers as previously described(48). Cells were centrifuged at 300 g, $4^{\circ}$C, 5 mins and the pellet was resuspended in HBSS and used for either whole node sequencing or macrophage isolation. CD11b positive macrophages were pulled down with Anti-mouse CD11b magnetic particles (BD Biosciences; Cat: 558013) according to the manufacturer's protocol. The isolated macrophages were resuspended in Trizol (Thermo Fisher Scientific) and stored at $-80^{\circ}$C till further analysis.

## Immunohistochemistry of frozen lymph node sections

One set of lymph nodes (both sacral and axillary) from both groups were snap-frozen in Tissue-Tek OCT (VWR, Radnor, PA) and sectioned at the Winship Cancer Institute's Pathology Core. Frozen sections of excised sacral and axillary nodes were blocked in 10% BSA in PBS and incubated with primary antibody overnight, and then secondary antibody for 2 h. Primary antibodies were anti-CD86 (Cat: MA1-10299), anti-iNOS(Cat: PA3-030A), anti-MHCII (Cat: MA5-16913) [all 3 from Thermo Fisher Scientific] and anti-IL12A (Acris Antibodies, San

Diego, CA; Cat: AM32704AF-N). These sections were detected using secondary antibodies conjugated with Alexa Fluor 647 or Alexa 680 (Thermo Fisher Scientific) and imaged by confocal microscopy using a Zeiss LSM 700.

### RNA Seq: Macrophage and whole node RNA isolation and library prep

RNA was isolated from macrophages or digested whole nodes stored in Trizol (ThermoFisher Scientific) per the manufacturer's instructions. Briefly, macrophages from control exosomes (n=2), pIC exosomes (n=2) and PBS dye (n=2) were homogenized for 20–30 sec with a rotor-stator homogenizer (Kimble Chase, Vineland, NJ). The sample lysates were then transferred to 2.0 ml Phase Lock Gels – Heavy (Eppendorf, Hamburg, Germany). Choloroform was added to each sample and centrifuged at 12,000 rcf for 10 min, and the upper aqueous layer was transferred to a new tube. The RNA was precipitated in isopropanol and washed in 75% ethanol according to the standard TRIzol protocol. The RNA pellet was then resuspended in RNase-free water (Ambion, Austin, TX) according to the manufacturer's instructions. RNA quantity and integrity were assessed by examining the relative intensity of 18s and 28s rRNA bands using an Agilent 2100 Bioanalyzer and RNA6000 Pico LabChip Kit (Agilent Technologies).

10 ng of Macrophage RNA was used as input to a Clontech Smart Seq v4 kit (Clontech labs, Mountain View, CA) to generate double stranded cDNA per the manufacturer's instructions. The cDNA was quantified using the Qubit HS DNA kit (Thermo Fisher Scientific) and 1 ng was used to prepare libraries using a Nextera XT DNA library preparation kit (Illumina, San Diego, CA) per the manufacturer's instructions.

250 ng of whole node RNA was used as input in a Truseq Stranded mRNA library prep kit (Illumina) to generate cDNA libraries. Both libraries were quantified using a Qubit Fluorimeter (Thermo Fisher Scientific) and multiplexed samples were run on a HiSeq 2500 instrument (Illumina). Each library was sequenced for 2×40 million of 100-nucleotide reads.

### Data analysis

RNA seq analysis for the three conditions, each with two replicates, was performed using raw reads (101bp, paired end) from an Illumina HiSeq 2000 machine. The reads were first aligned to the mouse reference genome (mm10, UCSC), using TopHat with default parameters. The transcripts were assembled from the aligned reads using Cufflinks, and the transcript abundance was calculated in terms of FPKM (fragments per kilobase of exon per million fragments mapped)(49). To compare the expression profile of the different samples, we used Cuffdiff, a differential expression analysis tool provided with the Cufflinks package. The results from Cuffdiff were used to plot the gene expression distribution graphs using a custom R script

The pathway analysis was performed using the GAGE RNA Seq workflow for pathway enrichment analysis(50). The –Log10(P-value) of the pathways of interest was plotted in R. Details of the analysis and the custom R scripts used to generate the figures are publicly available at the Github web site https://github.com/shashidhar22/macrophageRnaSeq. The data is publicly accessible in Sequence read archive (SRA) under the accession numbers: SRP074717 and SRP074576.

All graphs were generated on Prism 6 (GraphPad Software Inc, La Jolla, CA) and data is presented as mean ± standard deviation.

### Near infrared image analysis

Mice were imaged in the Near-infrared system and all transport metrics were calculated as previously described(18). Packet transport was calculated by integrating the fluorescence signal under the curve of each intensity spike. These intensity spikes have recently been shown to directly correlate with reductions in lymphatic diameter due to intrinsic lymphatic pumping(51).

## Acknowledgements

The authors wish to thank David Nelson for his initial ideas about endotoxin tolerance induction in distal cells. The authors also acknowledge Ms Shweta Biliya for running next-generation sequencing at the Georgia Tech High Throughoput DNA Sequencing Core at the Parker H. Petit Institute Bioioengineering and Bioscience.

# Author Contributions

S.S. designed the experimental approach, performed the experimental work, analysed the data, coordinated the project and wrote the manuscript. P.E.H. assisted in proximity ligation assays, S.R. assisted in macrophage RNA Seq analysis, J.M. assisted in whole node RNA Seq analysis, and M.S. assisted in qPCR. J.B.D. helped with the design of in vivo experiments, analying the kinetics of lymphatic transport of exosomes, and editing the manuscript. F.V. conceived the study, participated in the design of specific experiments and edited the manuscript.

# Competing Financial interests

The authors declare no competing financial interests.

Supporting information

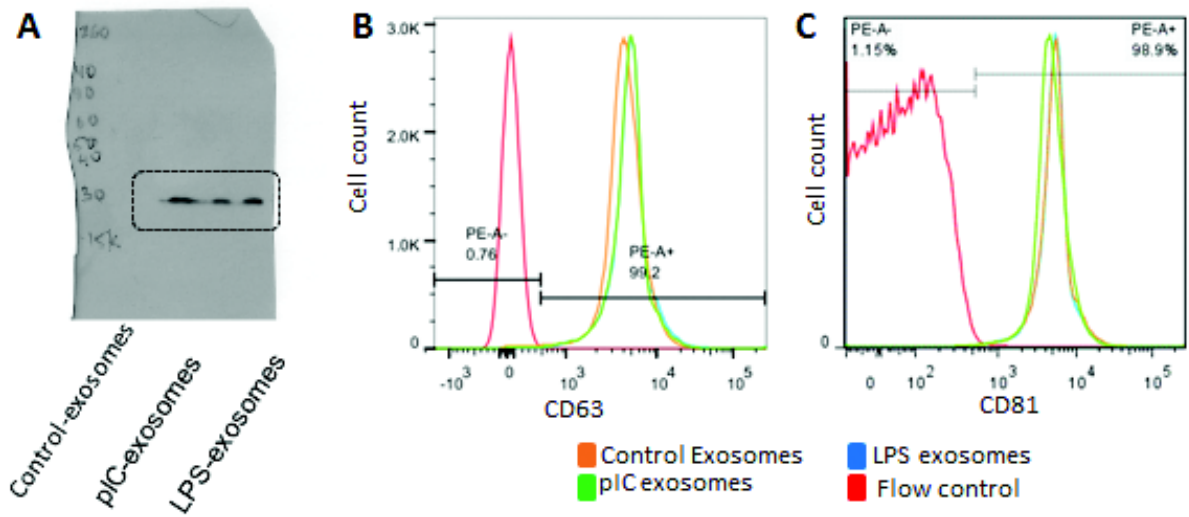

**Figure S1**: Characterization of exosomes used in the study. (**a**) Complete western blot of CD81 with control, pIC and LPS exosomes. Flow cytometry showing (**b**) CD63 and (**c**) CD81 levels on the exosomes.

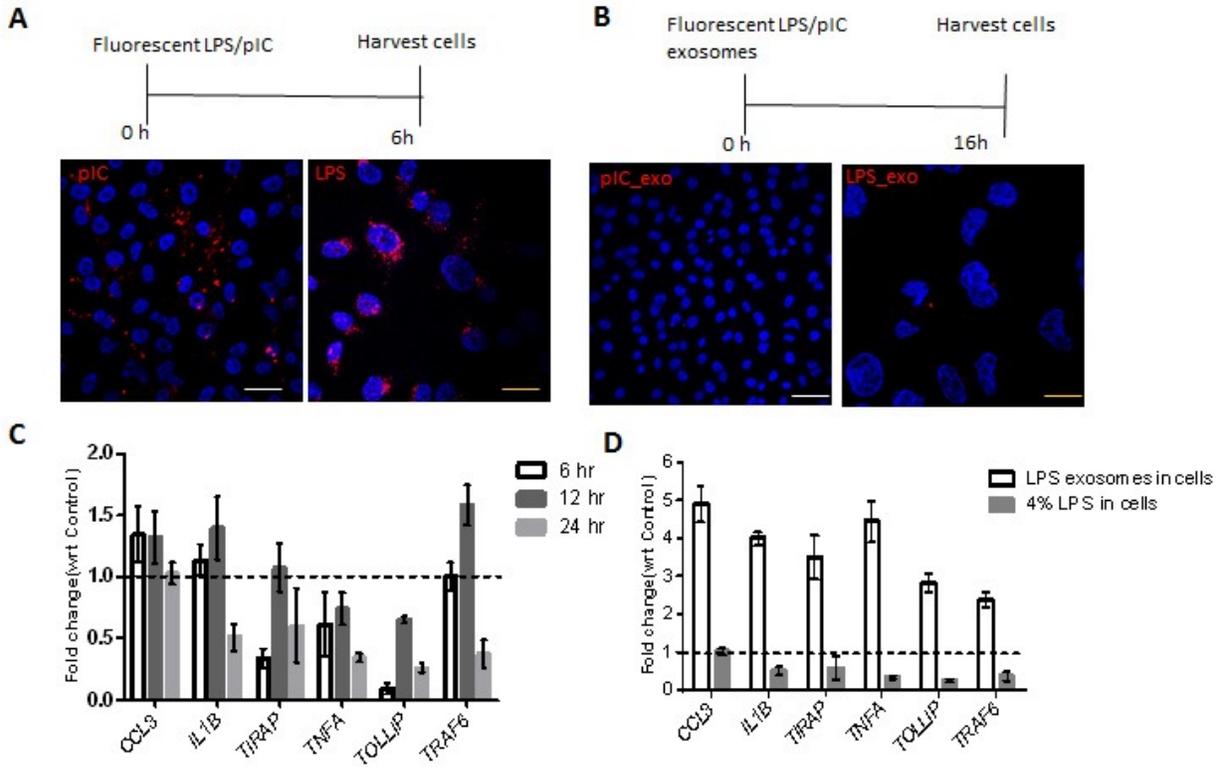

**Figure S2**: Estimating the carryover of TLR agonist from local cells to distal cells by exosomes. Confocal images showing (**a**) LPS-AF594 and pIC- Rhodamine uptake by parental cells and (**b**) exosomes from local cells treated with LPS-AF594 and pIC- Rhodamine added to distal cells to show no PIC and 4% LPS carryover. Scale bars, 50 μm. (**c**) Time course of gene expression in local cells after stimulation with 4% LPS and (**d**) Comparison of gene expression at 24 hours between local cells stimulated with 4% LPS and distal cells stimulated with LPS exosomes.

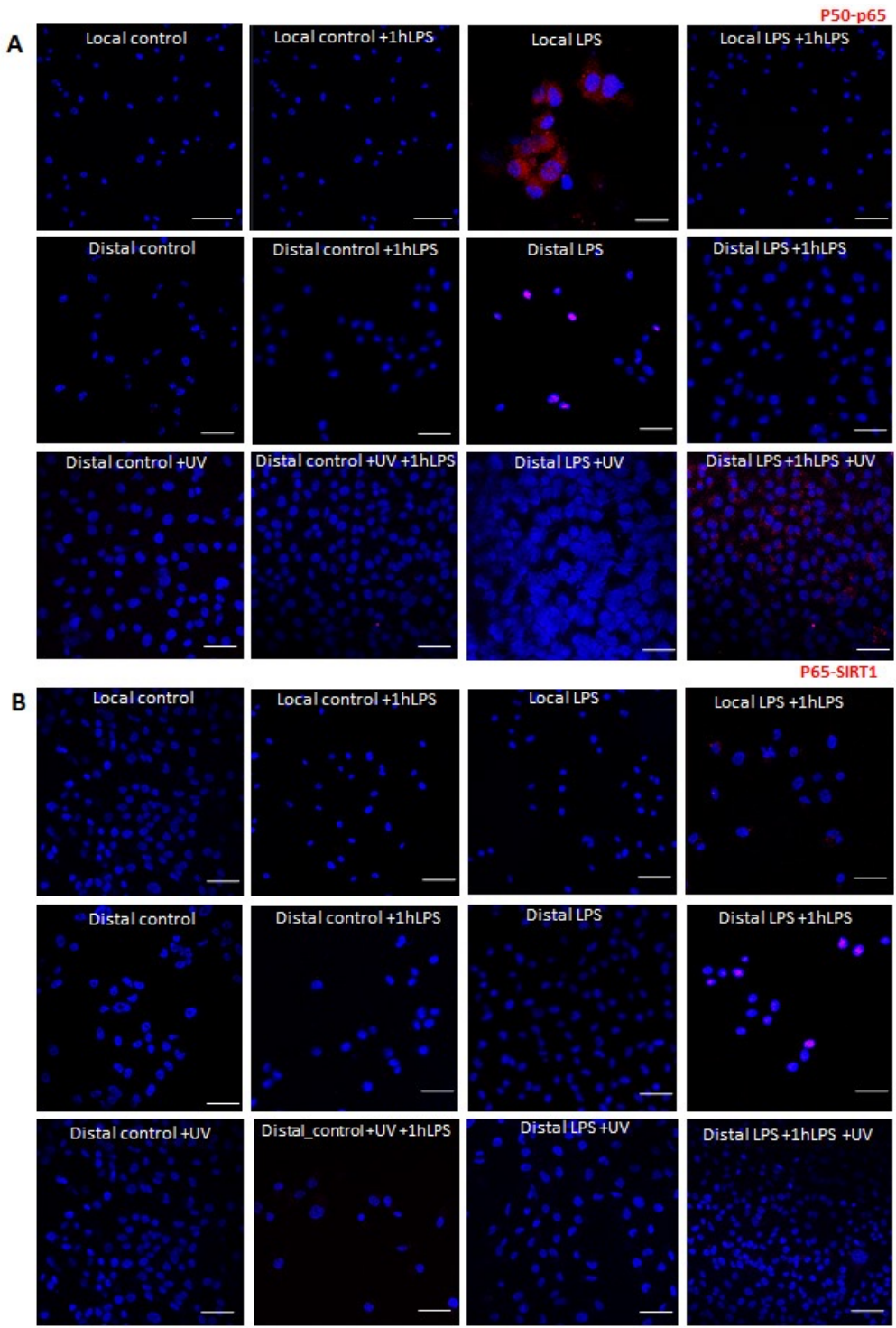

**Figure S3**: The LPS response in local and distal cells. Proximity ligation assay showing (**a**) the P50-P65 co-localization and (**b**) P65-SIRT1 in the cells indicated

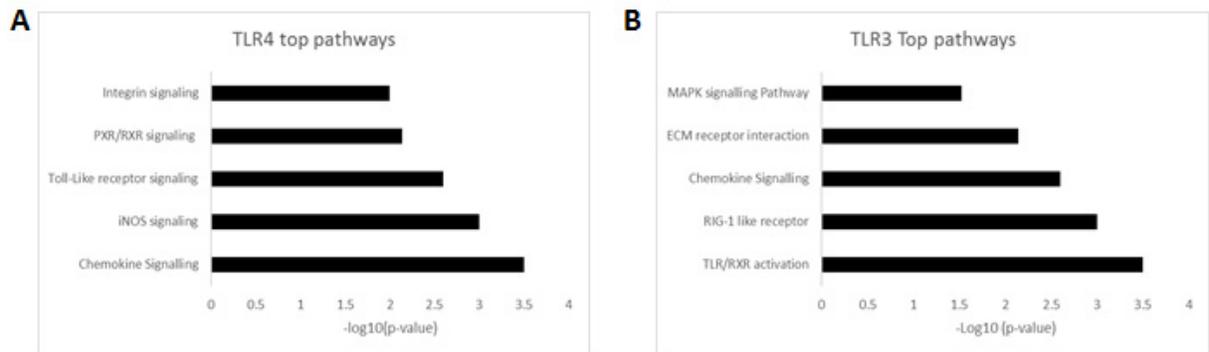

**Figure S4:** Pathways analysis of distal cells from microarray data. Pathways enriched in distal cells stimulated with (**a**) LPS exosomes and (**b**) pIC exosomes

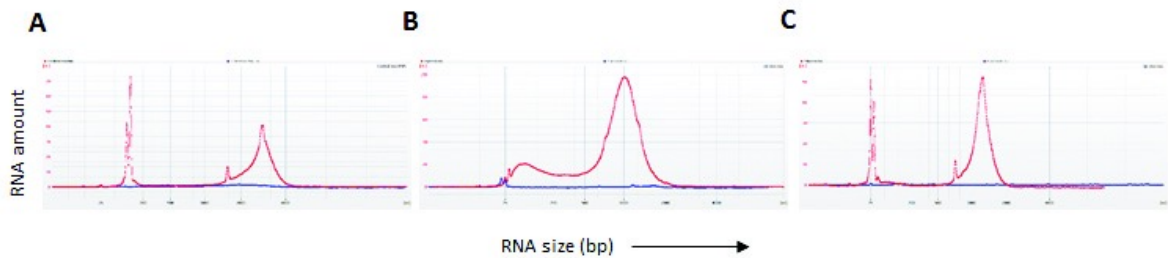

**Figure S5:** Effect of UV on nucleic acid content of exosomes. RNA size distribution profiles obtained on a Bioanalyzer pico RNA chip of (**a**) Control exosomes , (**b**) pIC exosomes,  and (**c**) LPS exosomes ; before (red lines) and after UV treatment (blue Line)

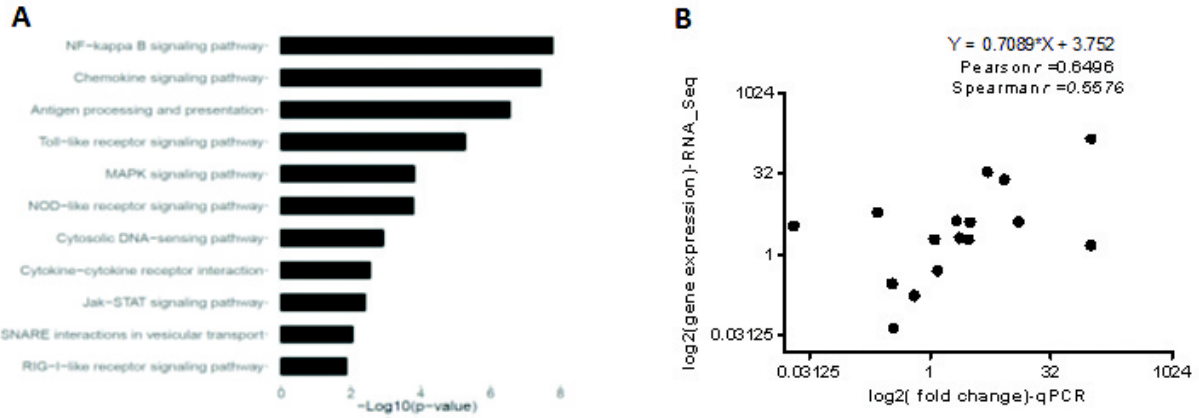

**Figure S6:** RNA-Seq of distal macrophages. (a) Pathways enriched in distal macropages with pIC exosomes as compared to PBS. (**b**) Scatter plots showing the correlation between the fold change detected via qCR when compared to RNA-Seq

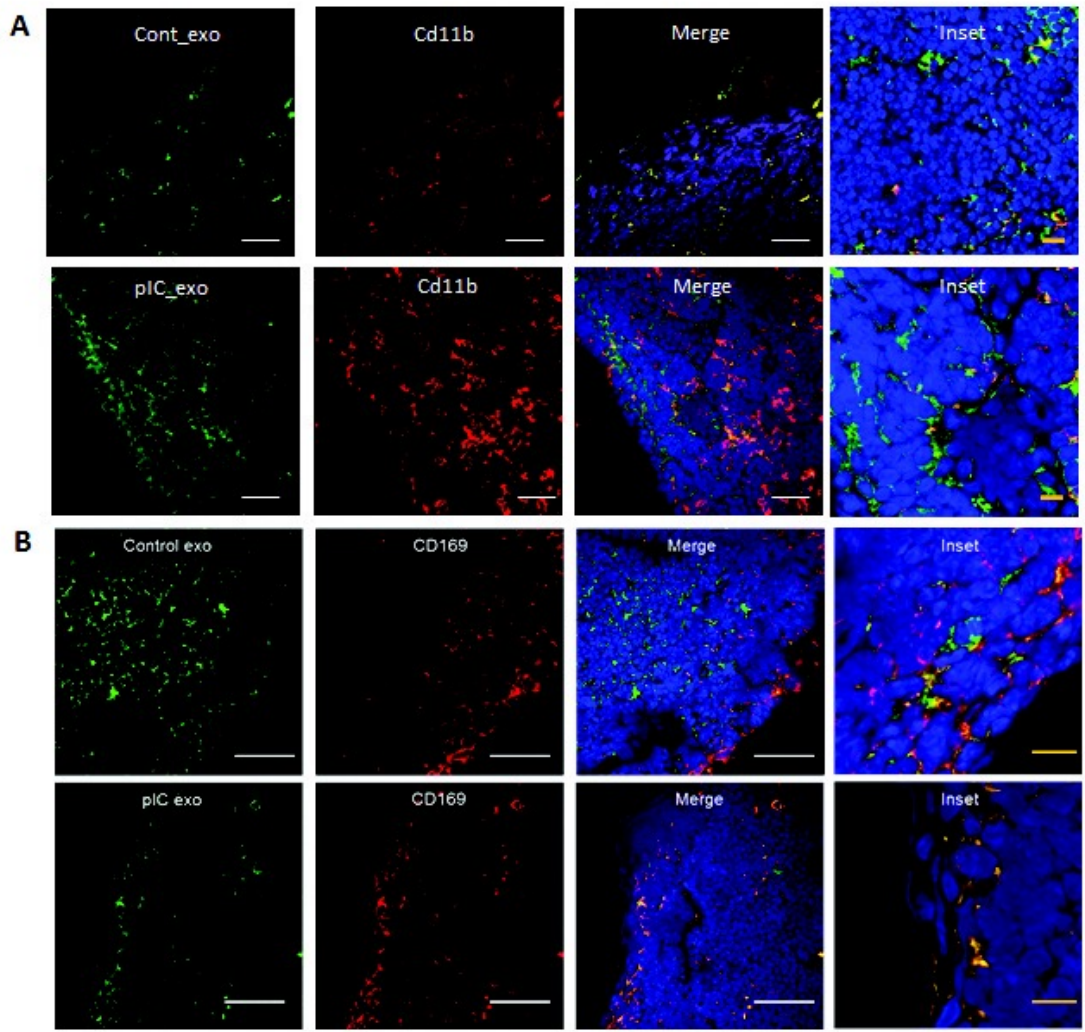

**Figure S7:** Macrophages retain both control and pIC exosomes **(a)** CD11b+ macrophages and (b) CD169+ subcapsular sinus macrophages retain control and pIC-exosomes.

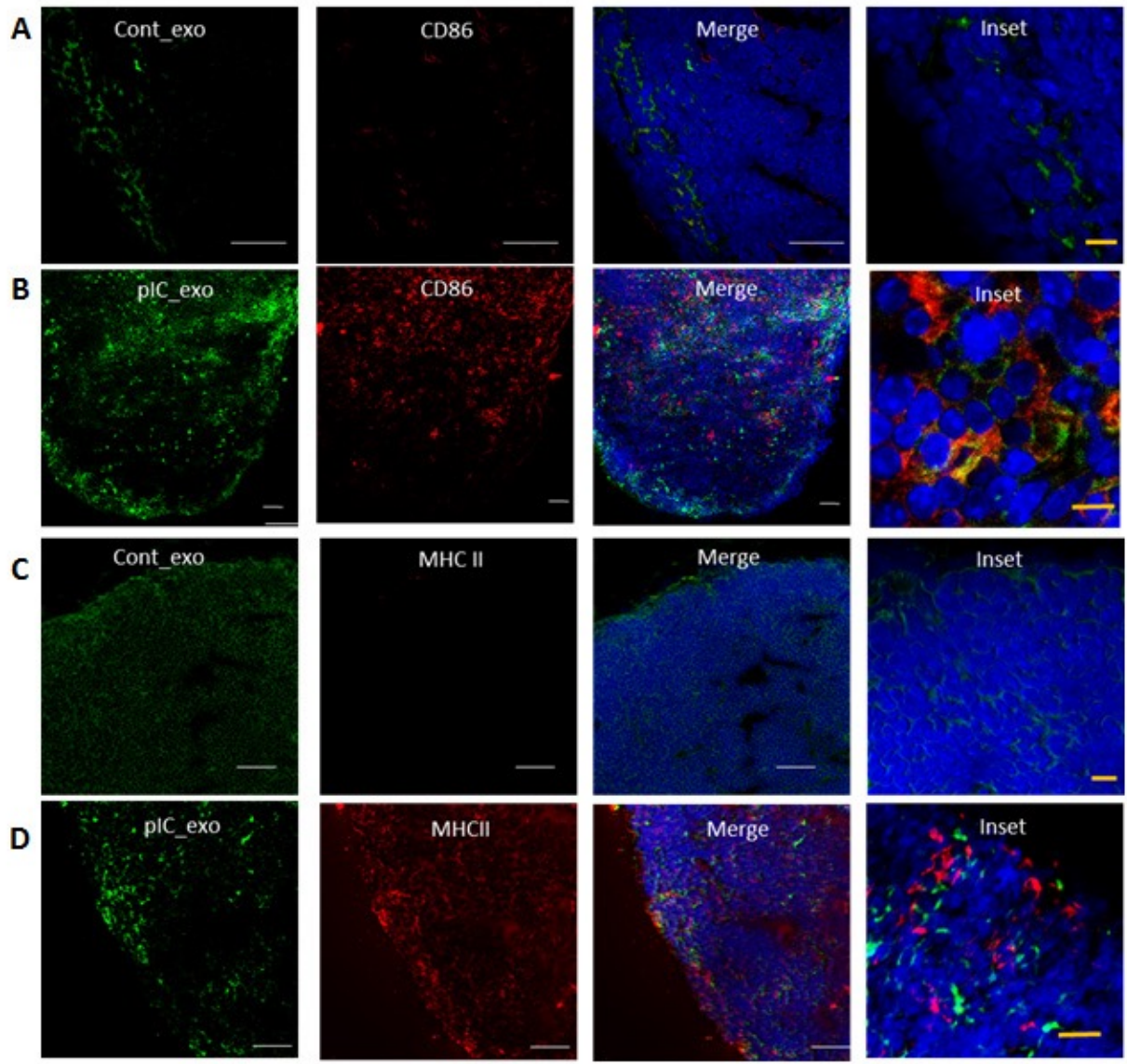

**Figure S8:** Validation of M1 markers Cd86 and MhcII expression in lymph node sections after exposure to control or pIC exosomes. White scale bars, 50 µm, yellow scale bars, 10 µm.

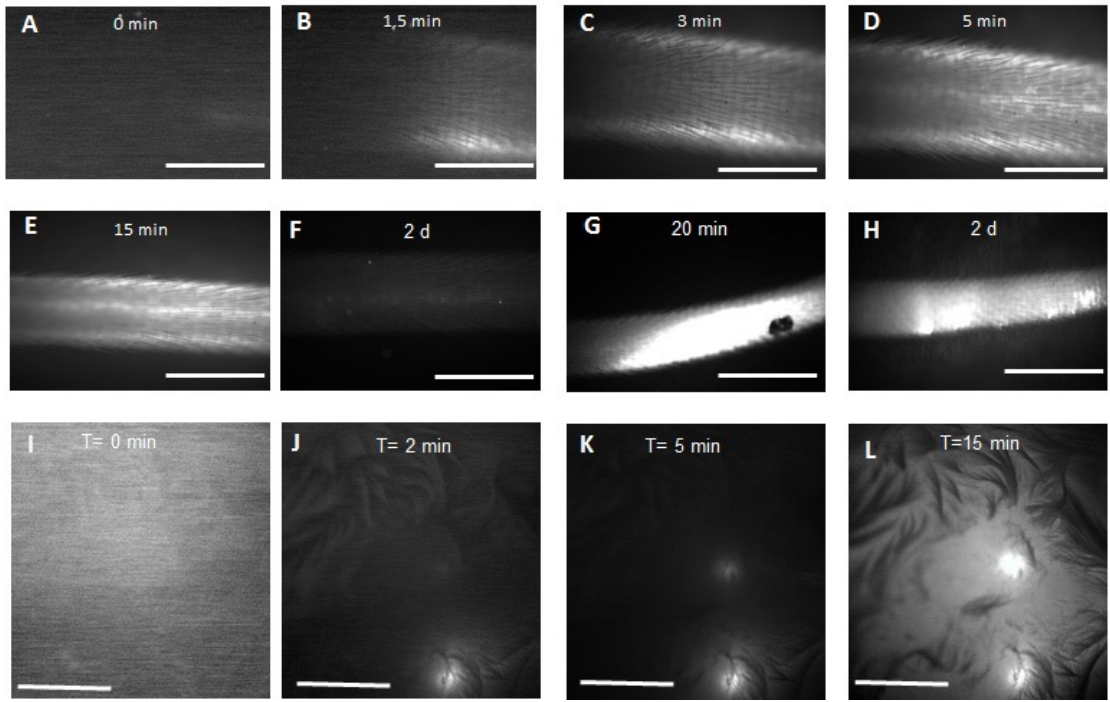

**Figure S9:** Lymphatic transport and retention of exosomes. pIC exosomes are seen in the lymphatic collecting vessels at (**a**) 0 mins, (**b**) 1.5 mins, (**c**) 3 mins, (**d**) 5mins, (**e**) 15 mins and (**f**) 2 days. The injection site is shown at (**g**) 20 mins and (**h**) 2 days. Exosomes are detected in the draining lymph node at (**i**) 0 mins, (**j**) 2mins, (**k**) 5 mins and (**l**) 15 mins+

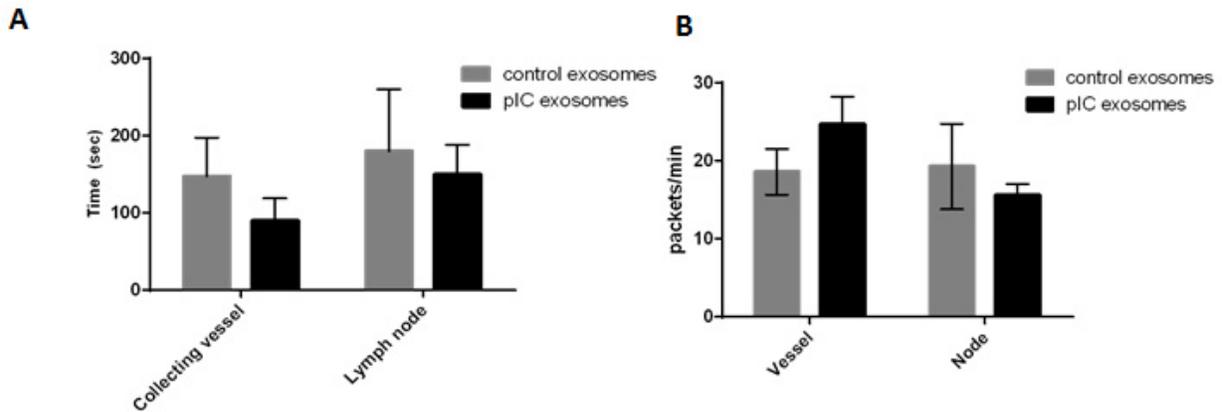

**Figure S10:** Kinetics of exosome transport in the lymphatics. (**a**) Arrival time of detectable levels of fluorescence for collecting vessels and draining lymph nodes. (**b**) Packet frequency of control and pIC exosomes in the collecting lymphatic vessels and nodes.

**Video S1**: Example video of pIC exosome arrival in the collecting vessels of a mouse 10 cm downstream from the site of intradermal injection. The dominant vessel is seen below and the non-dominant vessel is seen above. Video is played at 10X speed

**Video S2:** Example video of pIC exosome arrival in the draining (sciatic) lymph nodes of a mouse within minutes of intradermal exosome injection at the tip of the tail. The dominant node is seen below and the non-dominant vessel is seen above. Video is played at 10X speed

**Table S1:** List of human primers used in the study

**Table S2:** List of mouse primers used in the study

**Table S3:** Genes enriched in distal cells with LPS exosomes vs unstimulated HEY cells

**Table S4:** Genes enriched in distal cells with pic exosomes vs unstimulated HEY cells

**Table S5**: Genes enriched in pIC macrophages with respect to PBS macrophages

**Table S6:** Genes enriched in pIC macrophages with respect to control macrophages

**Table S7:** Genes enriched in pIC exosomes with respect to PBS in whole lymph nodes

**Table S8:** Genes enriched in pic exosomes vs control exosomes in whole node